\documentclass[12pt]{article}

\begin{document}

\begin{center}
  {\bf\Large Transition Radiation}
\end{center}
\begin{center}
{\bf\Large and the}\end{center}
\begin{center}
{\bf\Large Origin of Sonoluminescence}
\end{center}

\vspace{1cm}

\begin{center}
Bj\o rn Jensen\\{\it Faculty of Science,\\ Sogn and Fjordane
College,\\ N-6851 Sogndal, Norway.\\ {\tt
e-mail:bjorn.jensen@hisf.no}}
\end{center}
\begin{center}
Iver Brevik\\{\it Division of Applied Mechanics,\\ Norwegian
University of Science and Technology,\\ N-7491 Trondheim,
Norway.\\ {\tt e-mail:iver.h.brevik@mtf.ntnu.no} }
\end{center}

\vspace{.5cm}

\begin{center}
  (Revised version, February 2000. Typeset using Plain LaTex.)
 \end{center}


\begin{center}
{\bf Abstract}
\end{center}

 \begin{flushleft}
 \hspace{.5cm} It has been shown by Liberati et al.
 [quant-ph/9904013] that a
 dielectric medium with a time-dependent
 refractive index may produce photons.
 We point out that a free
 electric charge which interacts with such a medium will emit quantum
 mechanically modified transition radiation in which an
 arbitrary odd number of photons will be present. Excited atomic
 electrons will also exhibit a similarly modified emission
 spectrum. This effect may be directly observable in connection
 with sonoluminescence.
\end{flushleft}

\begin{flushleft}
{\bf\small PACS numbers: 12.20.Ds, 03.70.+k, 78.60.Mq, 42.50.Lc}
\end{flushleft}

\newpage

There exist mainly two ''schools'' in the effort to understand and
explain sonoluminescence. One school argues in favor of the view
that sonoluminescence can be understood on the basis of
gas-dynamics (see \cite{putterman}, e.g.). In the second main
approach it is argued that the flash of light which is emitted
from sonoluminescent bubbles is essentially due to energy which is
released from the quantum vacuum (see \cite{schwinger}, e.g.). In
this paper we will to some extent attempt to bridge the gap
between these apparently incompatible views.

Our starting point is the physical scenario for explaining
sonoluminescence which was presented in \cite{sciama}. There the
authors showed that sufficient energy in principle may be released
from the electro-magnetic vacuum in a dielectric medium with a
time-dependent refractive index so as to account for the radiation
emitted from sonoluminescent bubbles. The authors adopted a model in which there is
a {\it homogeneous} medium of time-varying refractive index. A detailed interaction mechanism, whereby this radiation can be released, was not specified. The approach taken in \cite{sciama} was thus purely phenomenological. We note that there are actually two kinds of phenomenology involved here: (i) already the use of the refractive index as such means that the complicated interaction between field and matter is described in terms of one single scalar parameter, the refractive index. It is this fact that gives rise to peculiar properties of phenomenological electromagnetic theory, such as the space-like nature of the total four-momentum of a radiation field within a medium (cf., for instance, \cite{brevik79}). We shall assume, such as in \cite{sciama}, that the use of the refractive index is meaningful in a quantum context also (otherwise, we would have to resort to some kind of many-body theory, which would be an extremely complicated approach). From a physical viewpoint one may interpret the refractive index as a time-dependent external field, producing pairs of photons. (ii) The second kind of phenomenology in \cite{sciama} is the assumption about homogeneity of the medium. Of course, this is a very rough model, but we agree with the authors in that it ought to be a reasonable first step towards a realistic theory. (In a later paper \cite{liberati99}, the same authors take finite volume effects into account.)

Characteristic for the above-mentioned approach is that the produced electromagnetic energy has to be calculated via the Bogolubov transformation \cite{birrell}. The calculational method is essentially the same as that found elsewhere when "sudden" changes are involved, as for instance in case of emitted radiation energy from the sudden production of a cosmic string \cite{parker87}.

Our main purpose with the present paper is to make a first step away from the phenomenological level and present a simple account of the photon production in terms of the excited or ionized atoms in the sonoluminescent bubble. This kind of approach may reveal properties of the emission spectrum that are connected with the interaction between vacuum and a time-dependent refractive index. That is, we may obtain some clues telling us about the relationship between the "hydrodynamic" and the "vacuum" interpretations of sonoluminescence. The present paper is to our knowledge the first attempt to bridge these two interpretations (although reference ought in this context be made to the very recent paper of Motyka and Sadzikowski \cite{motyka99} dealing with atomic collisions and sonoluminescence).

In this short paper we will show  that a free electron, which
interacts with a dielectric medium with a time-dependent
refractive index,  induces production of any odd number of
photons from the vacuum at the tree-level. The process can be seen
as a quantum mechanically (and non-perturbatively) modified
classical transition radiation process\footnote{Two particularly
nice references on classical aspects of transition radiation can
be found in \cite{ginzburg1,ginzburg2}.}. Excited atomic electrons
will also exhibit a similarly modified emission spectrum. We
suggest that these processes may provide partial mechanisms for
the observed emission of (at least some of the) photons from
sonoluminescent bubbles.

This brief communication is organized as follows. We will first
briefly consider the model which was presented in \cite{sciama}
with a special eye towards the question of a consistent
quantization of the electro-magnetic field in a medium with a
time-dependent refractive index. We then discuss qualitatively the
emission of photons from free and bound electrons which live in
such a medium. At the end of this paper we point out one
experimental consequence of our findings. We also point out a new
direction for further theoretical investigations into the question
of the possible importance of transition radiation in connection
with sonoluminescence.


We follow \cite{sciama} and choose to write Maxwell's equations in
a dielectric medium with a time- and position-independent
dielectric constant $\epsilon$, and with the magnetic permeability
$\mu$ set to unity, as (we set the speed of light, $\epsilon$ and
$\mu$ in the true vacuum to unity in the following)
\begin{eqnarray}
\mathrm{div}\,\mathbf{A}+\epsilon\,\partial_t\phi &=&0\, ,\\
-\epsilon\,\partial_t^2\phi +\mathrm{div}\,(\mathrm{grad}\,\phi )
&=&0\, ,\\
-\epsilon\,\partial_t^2\mathbf{A}+\mathrm{grad}\,(\mathrm{div}\,\mathbf{A})&=&0\,.
\end{eqnarray}
$\phi$ is the electro-magnetic scalar potential, and $\mathbf{A}$
is the three-vector potential. The first of these equations can be
seen as a ''generalized'' Lorentz-condition. We will further {\it
choose} potentials on the classical level such that
\begin{equation}
\phi =0 \,\, ,\,\, \mathrm{div}\,\mathbf{A}=0\, .
\end{equation}
The resulting wave equation for the electro-magnetic three-vector
field-potential then has solutions in the form
\begin{equation}
\mathbf{A}_{\mathbf{k}\lambda}=\mathrm{N}_\mathbf{k}\mathbf{e}_{\mathbf{k}\lambda}
e^{i\mathbf{k}\cdot\mathbf{x}-i\omega t}\,\, ,\,\,
\mathbf{k}\cdot\mathbf{A}_{\mathbf{k}\lambda}=0\, ,\, \lambda\in\{
0,1,2,3\}\, ,
\end{equation}
where ${\bf e}_{\mathbf{k}\lambda}$ is the polarization-vector,
$\mathrm{N}_{\mathbf{k}}$ is an normalization constant with
respect to some suitably defined inner product, and $\omega$ and
$\mathbf{k}$ are related by
\begin{equation}
-\epsilon\,\omega^2+\mathbf{k}^2=0\, .
\end{equation}
The polarization vectors can always be chosen such that
$\eta_{\alpha\beta}e^\alpha\, _{\mathbf{k}\lambda}e^\beta\,
_{\mathbf{k}\lambda '}=\eta_{\lambda \lambda '}$, where $\eta$
denotes the Minkowski metric with $\mathrm{sgn}(\eta )
=(-1,+1,+1,+1)$. We will let $\lambda =1,2$ refer to the
transverse components of the electro-magnetic field, while
$\lambda =0$ and $\lambda =3$ refer to the scalar and the
longitudinal components respectively.

It was assumed in \cite{sciama} that the dielectric constant of
the medium suddenly changes from its initial constant value, and
to another constant final value, i.e. it was assumed that
\begin{eqnarray}
&&\epsilon (t)
\rightarrow\epsilon_\mathrm{initial}=\mathrm{constant}\,\, ;\,\,
t\rightarrow -\infty\, ,\\ &&\epsilon (t)
\rightarrow\epsilon_\mathrm{final}\,\,\, =\mathrm{constant}\,\,
;\,\, t\rightarrow +\infty\, .
\end{eqnarray}
It was shown that such a change in the refractive properties of a
dielectric medium may give rise to production of photons. By
making the change in $\epsilon$ steep enough, it was shown that
sufficient energy in principle can be released so as to account
for the energy emitted in actual experiments with sonoluminescent
bubbles. The model does also reproduce the observed spectrum of
the photons which are emitted from such bubbles with a high degree
of accuracy.

There is at present a debate in the literature as to whether the {\it static} Casimir energy is large enough to account for the energy emitted by the sonoluminescing bubble. It was argued in \cite{iver} that the static Casimir energy is by far too small. It is only fair to mention, however,  that in a series of papers \cite{carlson} it has been argued that the static Casimir energy is large. As static considerations are not likely to be closely connected with the dynamic sonoluminescence problem, we will not pursue this subject further here.

We will now confront the question of whether the electro-magnetic
field can be quantized in a consistent fashion when one assumes
eq.(7-8) to hold true . Our main finding is that if one treats the
''generalized'' Lorentz-gauge condition along the same lines as in
the standard Gupta-Bleuer formalism, and along with the assumption
of a time-dependent dielectric constant in eq.(7-8), one gets the
conditions
\begin{equation}
\phi |\psi\rangle =0\, ,\, \mathbf{A}_{\|}|\psi\rangle =0\, ,
\end{equation}
{\it by necessity} at $t\rightarrow +\infty$, where
$\mathbf{A}_\|$ is the longitudinal part of the field-potential
($|\psi\rangle$ denotes any photon-state) {\it without the need to
invoke gauge-invariance}, in addition to the condition
\begin{equation}
\mathrm{div}\,\mathbf{A}_\bot |\psi\rangle =0\, ,
\end{equation}
where $\mathbf{A}_\bot$ is the transversal part of the
field-potential . Let us see how this comes about.

In the quasi-static asymptotic regions ($t\rightarrow\pm\infty$)
the electro-magnetic field can be quantized along the usual lines.
This implies the existence of two in general inequivalent
descriptions of the quantized field. In the in-region at infinite
past the field can be quantized as
\begin{equation}
A^{\mu\mathrm{in}}=\sum_{j\lambda}(
a^\mathrm{in}_{j\lambda}A^{\mu\mathrm{in}}_{j\lambda}+
a^{\mathrm{in}\dagger}_{j\lambda}A^{\mu\mathrm{in}*}_{j\lambda})\,
;\, \mu\in\{0,1,2,3\}\, ,
\end{equation}
where it is assumed a finite quantization volume. Note that this
discussion is a general one in which we do {\it not} impose the
constraints in eq.(4). The
$\{A^{\mu\mathrm{in}}_{j\lambda}\}$-modes represent a complete set
of solutions of the wave-equation with respect to some suitably
defined inner-product $(\, ,\, )$. $j$ refers to the $k_j$-mode. A
canonical (Heisenberg) vacuum state is as usual defined by
\begin{equation}
a_{j\lambda}^\mathrm{in}|0;\mathrm{in}\rangle =0\, .
\end{equation}
Similarly, in the infinite future we have
\begin{equation}
A^{\mu\mathrm{out}}=\sum_{j\lambda}(
a^\mathrm{out}_{j\lambda}A^{\mu\mathrm{out}}_{j\lambda}+
a^{\mathrm{out}\dagger}_{j\lambda}
A^{\mu\mathrm{out}*}_{j\lambda})\, ,
\end{equation}
which defines a vacuum state by
\begin{equation}
a_{j\lambda}^\mathrm{out}|0;\mathrm{out}\rangle =0\, .
\end{equation}
The descriptions of the electro-magnetic field in terms of the in-
and out-states are not in general physically equivalent, since the
products $\beta_{j'\lambda ';j\lambda}\equiv
(A^{\mu\mathrm{in}}_{j'\lambda
'},A^{\mu\mathrm{out}*}_{j\lambda})$ do not vanish in general.

Let us now impose the generalized Lorentz-gauge condition on the
positive frequency part of the operators above, i.e. consider the
general condition
\begin{equation}
(\partial_\nu A^{\nu
(+)}+\epsilon\,\partial_t\phi^{(+)})|\Psi\rangle =0\, ;\, \nu\in\{
1,2,3\}\, ,
\end{equation}
where $|\Psi\rangle$ is an arbitrary quantum state. $(+)$
indicates the positive frequency part of the operator. This
condition reduces to the usual Gupta-Bleuer condition when
$\epsilon =1$.
 In this true vacuum case the Gupta-Bleuer condition implies the relation
\begin{equation}
(a^\mathrm{in}_{j3}-a^\mathrm{in}_{j0})|\Psi\rangle =0\,
,\,\forall j.
\end{equation}
Let us follow the arguments in \cite{sciama}, and assume that the
system starts off in the true vacuum situation (
$\epsilon_\mathrm{initial}\equiv 1$, i.e.), and that the
refractive index is turned on after some time. At $t\rightarrow
+\infty$ the Gupta-Bleuer condition can naturally be written in
terms of the out-operators
\begin{equation}
(a^\mathrm{out}_{i3}-\sqrt{\epsilon_\mathrm{final}}
a^\mathrm{out}_{i0})|\Psi\rangle =0\, ,\,\forall i\, .
\end{equation}
The state-vector is unchanged, since it is time-independent in the
Heisenberg picture. The operator condition in this last expression
can be rewritten in terms of the operators appropriate for the
true vacuum situation. This results in the expression
\begin{equation}
(\sum_{j}\alpha_{ij}(a^\mathrm{in}_{j3}-\sqrt{\epsilon_\mathrm{final}}
a^\mathrm{in}_{j0})+
\sum_{j}\beta^*_{ij}(a^\mathrm{in}_{j3}-\sqrt{\epsilon_\mathrm{final}}
a^\mathrm{in}_{j0})^\dagger)|\Psi\rangle =0\, ,\, \forall i .
\end{equation}
$\alpha_{ij}$ and $\beta_{ij}$ are the Bogolubov-coefficients
which relate the in- and out-states (see \cite{birrell}, e.g.).
The condition in eq.(16) can be used to eliminate one of the
in-operators. We can then in particular re-express eq.(18) as
\begin{equation}
(1-\sqrt{\epsilon_\mathrm{final}})\sum_{j}(\alpha_{ij}a^\mathrm{in}_{j(0,3)}+
\beta^*_{ij}a^{\mathrm{in}\dagger}_{j(0,3)})|\Psi\rangle =0\,
,\,\forall i .
\end{equation}
$(0,3)$ indicates that this condition holds for the scalar and
longitudinal components separately. This condition can be further
simplified, by appealing to the definitions of the Bogolubov
coefficients, to read
\begin{equation}
(1-\sqrt{\epsilon_\mathrm{final}})a^\mathrm{out}_{i(0,3)}|\Psi\rangle
=0\, ,\forall i .\end{equation} Hence, consistency requires that
we quantize the physical, i.e. the transversal, degrees of freedom
in the out-region in the usual way, but put the scalar and the
longitudinal components {\it identically to zero}, as was
indicated in eq.(9). The physical content of the resulting quantum
theory is of course the one we would expect. However, what is
remarkable, and very unexpected, is that we arrived at our
conclusions eq.(9) and eq.(10) without the need to appeal to
gauge-invariance, or the equations of motion, at $t\rightarrow
+\infty$, as in the standard vacuum theory. It is unclear to us
what the deeper significance of this property is. We emphasize
that this feature is a general one, since the end-result is {\it
independent} of the details of how the dielectric constant
$\epsilon$ changes with time.

We also note another feature which deserves to be mentioned. The
$|\Psi\rangle$- state is constrained by two relations which
involve in-operators, namely the constraint in eq.(16) and the one
in eq.(19). Since the $a^\mathrm{in}_{j(0,3)}$-operators are
linearly independent, the condition in eq.(19) must hold for each
term in the sum separately, i.e.
\begin{eqnarray}
(\alpha_{ij}a^\mathrm{in}_{j(0,3)}+
\beta^*_{ij}a^{\mathrm{in}\dagger}_{j(0,3)})|\Psi\rangle =0\,
,\,\forall i,j .\end{eqnarray} It follows from these relations
that the unphysical sector of $|\Psi\rangle$,
$|\Psi\rangle^\mathrm{unphys.}$, can be represented as a coherent
state, i.e.
\begin{equation}
|\Psi\rangle^\mathrm{unphys.}=
\mathrm{N}e^{\sum_{ij}\Gamma_{ij}(a^{\mathrm{in}\dagger}_{j3}
a^{\mathrm{in}\dagger}_{j3}-
a^{\mathrm{in}\dagger}_{j0}a^{\mathrm{in}\dagger}_{j0})}
|0;\mathrm{in}\rangle\, .
\end{equation}
$\mathrm{N}$ is a non-vanishing normalization constant, and the
$\Gamma_{ij}$-coefficients are defined by
\begin{equation}
\Gamma_{ij}=-\frac{\beta^*_{ij}}{\alpha_{ij}}\, .
\end{equation}
When applied to this coherent state the condition in eq.(16)
implies
\begin{equation}
(a^{\mathrm{in}\dagger}_{j3}-a^{\mathrm{in}\dagger}_{j0})
|\Psi\rangle^\mathrm{unphys.}=0\, ,\, \forall j .\end{equation}
What all this amounts to is that the initial configuration of
scalar- and longitudinal excitations must have a particular form.
This form is dictated by the behaviour of the dielectric constant
in the future. We emphasize that this point is important, since
there was no guarantee {\it a priori} that the initial
photon-state does not contain physical scalar and longitudinal
excitations as they are defined in the future. However, the model
considered is consistent, at least at this level, since we can
always choose the unphysical sector to have the form in eq.(22).
Let us now turn to the question of the nature of the emission
spectrum from free and bound electrons which live in a medium with
a time-dependent refractive index.

As mentioned above, in \cite{sciama} the dielectric constant was
assumed to be a position-independent parameter, enabling one to
obtain a rough order of magnitude of the energy emission from the
medium. In order to actually understand what triggers the emission
of photons, one must look at the microscopic dynamics more
closely. There are two different physical mechanisms that are very
natural to consider: first, the molecules in the bubble may be
excited, eventually ionized, by photonic interaction. Secondly,
the molecules may experience this effect because of fluiddynamical
forces, induced by  shock waves. As for the latter possibility,
the fluid dynamical theory of sonoluminescence given by Kwak et.
al. \cite{kwak96}, based upon the theory of dense gas in the
bubble as given, in particular, by Wu and Roberts \cite{wu96},
   is quite impressive. Now, it is also an experimental fact that
   some amount of noble gases is required in the bubble to make the
    sonoluminescent process work. What is the physical reason for this?
    The explanation is as yet not completely clear, but is most probably
    related to the fact that that the spherical symmetry of the noble
    gas atoms prevent the excitation energy to become distributed over
    rotational or vibrational modes. The rotational symmetry thus helps
    to make the emitted frequences high.

From these physical considerations we adopt henceforth the following model: there exists a thin gas of free electrons in the bubble; we do not specify whether they are the result of photoionization, shock wave effects, or a combination of these. Mathematically, we take the initial quantum state at sufficiently early times to be
\begin{equation}
|i\rangle\equiv |1;\mathrm{in}\rangle\otimes
|0;\mathrm{in}\rangle\, ,
\end{equation}
where we have suppressed all quantum numbers. The first vector in
the tensor product describes the number of initial electrons,
while the second vector designates the number of initial photons.
We write the final state in the same vain as
\begin{equation}
|f\rangle\equiv |1';\mathrm{out}\rangle\otimes
|n;\mathrm{out}\rangle\, ,
\end{equation}
where we have put a prime on the fermionic state in order to
distinguish it further from the initial fermionic state. The
interaction term between the electron-field and the
electro-magnetic field in the bulk is
\begin{equation}
\mathcal{H}_\mathrm{I}= A_\mu j^\mu
=eA_\mu\overline{\psi}\gamma^\mu\psi\, ,\end{equation} where $e$
is the fundamental electric charge. $\psi$ denotes the
electron-positron field, $\overline{\psi}$ its conjugate and
$\gamma^\mu$ represent the usual gamma-matrices. This interaction
is the only one we will consider. Hence, we neglect any effects
which are associated with having to deal with a finite volume, or
effects which appear due to the motion of the surface of the
confining volume. However, we will briefly return to these issues
at the end of the paper.

The asymptotic properties of the dielectric constant is
characterized by eq.(7-8). The incoming electro-magnetic vacuum
state, as it is canonically defined at $t\rightarrow -\infty$, is
in general related to the outgoing electro-magnetic vacuum state,
as it is defined at $t\rightarrow +\infty$, by (see eq.(7.9) in
\cite{fulling} for the case when the $\alpha$- and
$\beta$-matrices are diagonal. It is straightforward to generalize
that expression to the one in eq.(28))
\begin{equation}
(\phi_0)^{-1}|0;\mathrm{in}\rangle=e^{(\frac{1}{2}\sum_{ij\lambda}\gamma^*_{ij\lambda}
a^{\mathrm{out}\dagger}_{j\lambda}
a^{\mathrm{out}\dagger}_{-j\lambda})}|0;\mathrm{out}\rangle\, .
\end{equation}
$\phi_0$ is in general an arbitrary constant, which we will set
equal to unity, $*$ denotes complex-conjugation, and
$\gamma_{ij}\equiv (\beta_{ij}/\alpha_{ij})^*$. This relation
follows due to the completeness of the Fock-spaces. From the
expression for $|0;\mathrm{in}\rangle$ in eq.(28) we see in
particular that any energy emission (in the form of photons as
they are defined in the future) from $|0;\mathrm{in}\rangle$ in
the far future must involve correlated pairs of photons. In
dielectric media it is only the electro-magnetic fluctuations
which are changed when one compares with the pure vacuum case, i.e.
when considered as free fields it is only the second quantization
of the electro-magnetic field which is modified when $\epsilon
>1$, while the second quantization of the electron-positron field
is identical to the one in the true vacuum. Hence, the part of the
Fock-space which is associated with the incoming electron in
eq.(25) is not altered during the change in the refractive
properties of the medium. It is thus appropriate to change the
notation in the definition of $|f\rangle$ into
\begin{equation}
|f\rangle =|1';\mathrm{in}\rangle\otimes|n;\mathrm{out}\rangle\, .
\end{equation}

Photon emission from a free electron at the tree-level in the
out-region is described by the S-matrix element
\begin{equation}
S_{fi}\equiv \int\langle f|\mathcal{H}_\mathrm{I}|i\rangle\, ,
\end{equation}
where $\int$ indicates integration over all space-time. By keeping
just the first non-trivial term in eq.(28) the S-matrix element
above can be expanded as

\newpage

\begin{eqnarray}
S_{fi}&=&\int \langle
f|\mathcal{H}_\mathrm{I}|1';\mathrm{in}\rangle\otimes
|0;\mathrm{out}\rangle+\nonumber\\ &+&\frac{1}{2}\int\langle
f|\mathcal{H}_\mathrm{I}(\sum_{ij\lambda}\gamma_{ij}^*a_{j\lambda}^\dagger
a_{-j\lambda}^\dagger
)|1';\mathrm{in}\rangle\otimes|0;\mathrm{out}\rangle+....\nonumber\\
&\equiv& S_{fi(1)}+S_{fi(2)}+....\, .
\end{eqnarray}
Since the electron-positron field is insensitive to a changing
refractive index we will neglect this part in the following in
order to simplify our formulas. At $t\rightarrow -\infty$ the $\{
A^\mu\}$-potentials have the general form in eq.(11). These
potentials are propagated forward in time by the equations of
motion which takes an explicitly time-dependent refractive index
into account (these equations are reproduced in \cite{sciama}). In
order to compute the S-matrix we utilize the relation
\begin{eqnarray}
A^\mu &=& \sum_{j\lambda}(
a^\mathrm{in}_{j\lambda}A^{\mu\mathrm{in}}_{j\lambda} +
 a^{\mathrm{in}\dagger}_{j\lambda}A^{\mu\mathrm{in}*}_{j\lambda}) =\nonumber\\
 &=&\sum_{ij\lambda}( (\alpha_{ij}
 A^{\mu\mathrm{in}}_{j\lambda}+(\beta_{ij}
  A^{\mu\mathrm{in}}_{j\lambda})^*) a^\mathrm{out}_{j\lambda}+
 (\beta_{ij}  A^{\mu\mathrm{in}}_{j\lambda}+(\alpha_{ij}
  A^{\mu\mathrm{in}}_{j\lambda})^*)a^{\mathrm{out}\dagger}_{j\lambda})
 \end{eqnarray}
 where we in the last line have expressed the
 in-operators in terms of the out-operators. The
 $A^{\mu\mathrm{in}}_{j\lambda}$-modes have been calculated for a particular
 case in \cite{sciama} in terms of hyper-geometric functions.
 The relevant part of the S-matrix which describes
 emission of a single photon from an electron is then given by
 \begin{eqnarray}
 S_{fi(1)}=\int (\langle\mathrm{out};1|\sum_{ij\lambda}(\beta_{ij}
 A^{\mu\mathrm{in}}_{j\lambda}+(\alpha_{ij}
 A^{\mu\mathrm{in}}_{j\lambda})^*)a^{\mathrm{out}\dagger}_{j\lambda}
 |0;\mathrm{out}\rangle\otimes\mathrm{Fermions}_\mu )\, .\end{eqnarray}
 Such emission at the tree-level is in the literature called
 transition-radiation and is a well known phenomenon, although our quantum field
 theoretical treatment of it seems to be among the first ones.
 After $S_{fi(1)}$ has been computed it is  straightforward to
 compute $S_{fi(2)}$ which describes the emission of three photons
 of which two are back to back. When still further terms in eq.(28)
 are taken into account we find that an electron may emit any odd
 number of photons.

Sonoluminescent bubbles may consist entirely of noble-gas atoms
\cite{Lohse}. Such atoms have ionization energies of the order of
$15\, \mathrm{eV}$. As the observed radiation in experiments
with sonoluminescent bubbles have a typical energy of the order of
$3\, \mathrm{eV}\,-4\,\mathrm{eV}$, it can be argued
(see \cite{sciama}, e.g.) that the presence of free charges
in a sonoluminescent bubble is an unlikely hypothesis.
However, in practice, with a mixture of gases in the bubble, the complexity of the interactions (photonic or hydrodynamic) makes it conceivable that there exists a thin electron gas there nevertheless. And, as mentioned, this is at the core of our model.

It was argued in \cite{Belgiorno} that it is in principle possible
to experimentally verify whether the vacuum picture of
sonoluminescence is a viable one by measuring two-photon
correlations. Our main aim in this paper is to investigate whether
the inclusion of free or bound charges will induce further
properties to the radiation which is emitted by such bubbles.
Transition-radiation (meaning the emission of a {\it single}
photon from a charge) is a well known physical effect which will
be present in {\it any} dielectric medium with a time-dependent
refractive index as soon as free electric charges are present. On
the assumption of the presence of a thin gas of electrons inside
sonoluminescent bubbles it follows that transition radiation must
also be taken into account in order to understand
sonoluminescence. In this paper we have shown that a quantum
treatment of transition radiation reveals that a free electron
does not necessarily emit only a single photon, but it may in fact
emit any odd number of photons. This finding opens up new
directions for further experimental work. Clearly, since electrons
can emit any odd number of photons it will be induced correlations
between odd number of photons when one measures the {\em arrival
times} of photons from sonoluminescent bubbles.

It may seem straightforward, and therefore tempting,
 to compute
the efficiency of the production of the extra correlated
photon-pairs in transition radiation processes by putting the
Bogolubov coefficients which was computed in \cite{sciama} into
our formulas. We want to emphasize that the inclusion of the
accelerated cavity walls may be of {\em crucial importance} for
such estimates. However, transition radiation in connection with
accelerated layer surfaces appears to be a completely uncharted
area in the literature. Hence, realistic estimates of the
efficiency of the transition radiation mechanism to produce
correlated pairs of photons must await future studies of this
particular problem \cite{bjorn}.

\end{document}